\documentstyle[aps,prl,multicol,epsf]{revtex}

\begin{document}

\title{Scaling of avalanche queues in directed dissipative sandpiles}

\author{Bosiljka Tadi\'c$^{1,*}$ and   Vyatcheslav Priezzhev$^{2,**}$
}

\address{$^1$Jo\v{z}ef Stefan Institute,
P.O. Box 3000, 1001 Ljubljana, Slovenia \\
$^2$Bogolubov Laboratory of Theoretical Physics,
Joint Institute of Nuclear Research, 141980 Dubna, Russia }


\maketitle
\begin{abstract}
\newline
Using  numerical simulations and analytical
methods we study a two-dimensional directed sandpile automaton with
nonconservative random defects (concentration $c$) and varying
driving rate $r$. The automaton is driven
only at the top row and driving rate is measured by the number of
added particles per time step of avalanche evolution. The probability
distribution of duration of elementary  avalanches at zero driving rate
is exactly given by  $P_1(t,c) = t^{-3/2}\exp{[t\ln(1-c)]}$.
 For driving rates in the interval $0< r \le 1$
the avalanches are queuing one after another, making  increase
the periods of  non-interrupted activity of the automaton.
Recognizing the probability $P_1$ as a distribution of service time of
jobs arriving at a server with frequency $r$, the model
 represents a new example of the $\langle E,1,GI/\infty/1\rangle $ server
queue  in the queue theory. We study scaling properties of the
busy period and dissipated energy of sequences of non-interrupted
activity. In the limit $c \to 0$ and varying linear system size
$L \ll 1/c$ we find that at driving rates $r\le L^{-1/2}$
the distributions of duration and energy of the avalanche queues are
characterized by a multifractal scaling and we determine the
corresponding spectral functions. For $L\gg 1/c$
increasing of the driving rate somewhat compensates the energy losses
at defects  above the line $r \sim \sqrt{c}$. The scaling  exponents of
the  distributions in this region of phase diagram vary approximately
linearly with the driving rate.  Using properties of recurrent states
and the probability theory we determine analytically the exact upper
bound of the probability distribution of busy periods.
 In the case of conservative dynamics $c=0$ the probability
of a continuous flow increases as $F(\infty ) \sim r^2$ for
small driving rates.

\end{abstract}

\begin{multicols}{2}

\newpage

\section{Introduction}
In the past decade the sandpile type of cellular automata played
a special role in understanding of self-organized criticality in
nonlinear dynamical systems (for a recent review see Ref.\ \cite{DD-rev}).
In sandpile automata the properties of the dynamics
which are essential for the occurrence of self-organized critical states
can be monitored in a  direct manner. Apart from the relaxation rules,
these are the following properties: type of driving and time-scale
separation; conservation law of the dynamics; direction of mass flow;
role of boundaries. In addition, the Abelian nature of the toppling
rules in some sandpile automata enabled derivation of certain exact
results \cite{DR,exact}, in contrast to other dynamical  systems where
such calculations are not available.
Numerous  sandpile models, both deterministic and stochastic \cite{DD-rev}
are shown to exhibit dynamic critical states in the limit of
``infinitely slow'' driving (i.e., at zero driving rate $r=0$). In this
limit a new avalanche is initiated
only after the previous one has stopped, thus the time scale
separation is exactly observed. On the time scale of perturbations,
the avalanche evolution is seen as occurring instantly.
The existence of the critical states in the case of directed Abelian
sandpile automaton at zero driving rate has been proved exactly  by Dhar
and Ramaswamy \cite{DR}. At this point it is interesting to mention that the
model studied in Ref.\ \cite{DR} is characterized by  local driving and
deterministic conservative dynamics. The automaton with conservative
 stochastic dynamics, on the other hand,  has been shown
to belong to another universality class \cite{TD}. The presence
of nonconservative defects in Dhar-Ramaswamy automaton leads to a
subcritical behavior \cite{Tetal}.

The behavior of driven dynamical systems at finite driving rates
($r>0$) represents an important subject both for theoretical and
practical reasons. A finite driving rate may appear either as a
control parameter set from outside, or as a
 probability distribution originating from another coupled stochastic
process.
In practice, the systems are driven by an external field, which oscillates
with a finite frequency. Examples are Barkhausen noise \cite{BN-r},
integrate and fire oscillators \cite{IFO}, granular material
in rotating drums \cite{drums}, etc. Queuing jobs at a server
\cite{queue-books}, e.g., in teletraffic, is an example where the
frequency of arriving jobs is given by a random process.

Theoretically at finite driving rates $r>0$  the probability that
 a new avalanche starts before previous one has stopped increases with
increasing $r$. This obviously
leads to different statistics of avalanches, where an avalanche is
understood to represent a  non-interrupted activity of the system.
For large driving rates
a continuous flow (an avalanche which never stops) is expected in sandpiles.
Similarly, a single spanning cluster may occur in driven disordered systems.
Therefore, a time scale separation becomes
less and less apparent with increasing $r$. In addition, by
increasing  driving rates, the local driving looses its strict sense.
Thus fast driven sandpiles are placed between strictly local driving,
where the system is driven at a single (random) site, and global driving,
where the same perturbation applies to all sites in the system.
The role of the conservation law (conservation of number of grains in the
interior of the system) is also expected to be changed at finite
driving rates. In $r\to 0$ limit, locally driven nonconservative systems
appear to be subcritical \cite{noncons-local}, whereas when the driving
is global the critical states may appear even if the dynamics is
dissipative \cite{noncons-global,IFO}.
So far neither an exact theory nor a renormalization-group analysis of
fast driven critical systems has been done. A general questions as the
existence of critical states at finite driving rates and   universal
scaling  properties of the system in the limit of large distances and
long times remain yet to be understood.

Recently two numerical simulations elucidated certain important
properties of sandpiles at finite driving rates.
In a $1$-dimensional ricepile model Corral and Paczuski \cite{Al-Maya}
have first introduced avalanches for a finite driving rate as
non-interrupted periods of activity and have
shown that these avalanches diverge  at  rates $r\ge r_c(L)
\sim L^{-0.20}$ for a given  $L$.
The rational behind this conclusion is that an ever-running avalanche
occurs for driving rates $r\sim 1/\langle t\rangle _0$, where
$\langle t\rangle _0 \sim L^{z(2-\tau _t)}$ is the average duration of
avalanches in zero driving rate \cite{Al-Maya}.
 In another example  Barrat {\it et al.} \cite{BTW-r} have shown
that in a $2$-dimensional symmetric
Abelian sandpile model mixing of time scales at finite driving rates
leads to correlations which appear to violate the fluctuation-dissipation
theorem.

In this work we study a simple $2$-dimensional  model with strictly
directed flow of grains and deterministic toppling rules.
We add particles {\it only at the top row}
 with driving rate $r$. The driving rate  $r$ is defined as a
number of added particles per time step of avalanche evolution.
We consider both conservative and nonconservative dynamics.
A fraction of sites $c$ are considered as annealed nonconservative
defects. By toppling at a defect site two grains are lost, thus
affecting the propagation of avalanche below that site \cite{comment-sb}.
When $c=0$ the dynamics is strictly conservative.
In the $r=0$ limit and $c=0$ the model
has been exactly solved by Dhar and Ramaswamy \cite{DR}. The critical
states where shown to consist of heights
$h=0$ and $h=1$ occurring  with equal probability 1/2 at each lattice site.
The duration of an avalanche is given by the probability distribution
for large $t$
\begin{equation}
P_1(t) \sim t^{-3/2} \ .
\label{p1_0}
\end{equation}
In the presence of nonconservative defects it has been shown \cite{Tetal,JT}
that the screening of the power-law distribution in Eq.\ (\ref{p1_0}) occurs
as
 \begin{equation}
P_1(t) \sim t^{-3/2}\exp({-t/\xi }) \ ;\   \xi \sim 1/c \ .
\label{p1_c}
\end{equation}
In this paper we will refer to the distributions in Eqs.\ (\ref{p1_0})
and (\ref{p1_c}) as
probability distributions of {\it elementary} avalanches, to be
distinguished from the {\it combined} avalanches, which
occur at finite driving rates and which consist of  series of elementary
avalanches.
Up to relatively high driving rates $r=1$ the model has the property
that successive elementary avalanches are running one after the other,
in contrast to cases studied in Refs.\ \cite{Al-Maya,BTW-r}, where merging
of avalanches may occur at any finite $r$.
After an elementary avalanche is over  the system is characterized
by statistically unchanged distribution of heights, owing to a weak
correlation between the avalanches in the recurrent states.
A finite probability of avalanche collision, which accelerates flow
of grains, occurs in this model only for $r>1$. Here we restrict the
study to the case
$r\leq 1$, where the formation of avalanche queues is a dominant
phenomenon which determines the scaling properties of the system.

The problem of avalanche queue in our model can be regarded as an
example of the server queue \cite{queue-books} of the class
$\langle E,1,GI/\infty,1 \rangle$, which is studied as a model in the
analysis of various applied problems, e.g., in telecommunications,
insurance, etc.  These analogies are made clear by
noticing that the avalanches of a directed sandpile model have
the corresponding terms in the language of the queue theory as
follows:  (i) elementary avalanche --- customer;
(ii) duration of elementary avalanche --- service time;
(iii) driving rate --- frequency of arrivals of customers;
(iv) number of elementary avalanches coexisting at a given moment
of time --- number of working servers;
(v) duration of a combined avalanche --- busy period.
The notation $\langle E,1,GI/\infty,1 \rangle$ means that we
 deal with customers arriving by one and being served by one. The letter
$E$ means that the arrival times are generated by a Bernoulli process
with the distribution
$Prob(t=k) = p^k(1-p); k=0,1,2,... ; p>0$\ .
The symbol $GI/\infty$ means that the service times are
identical independently distributed (i.i.d.) random variables and the
infinite number of
servers provides  a non-restricted number of customers which can be served
simultaneously. Despite a huge literature devoted to  this subject
\cite{queue-books}, most of papers
focus on the distribution of $q_n$, the number of working servers at a
given moment of time.  The importance of the tail behavior of the busy
period distribution for fluid queues in telecommunications, generalized
processor sharing and other applications was recently pointed out
in Ref.\ \cite{Zwart}.
Here we concentrate on some other properties of
the queue: the scaling behavior  of the busy period and dissipated  energy
distribution.

Given the duration of elementary avalanches in Eq.\ (\ref{p1_0}), we may
conclude that the directed sandpile model for $r\leq 1$
represents a special case of the queue theory with the power-law
distribution of the service time  with the exponent
$\nu =3/2 -1 =1/2$.  This implies
that the average service time per customer diverges $\langle t\rangle \to
\infty $. In practice, service times are restricted to finite values,
which corresponds to the power-law distribution  with $1 <\nu <2$
\cite{Zwart} . This may explain  why the queue
with  the distribution of the type given in Eq.\ (\ref{p1_0}) has not
been studied so far.
In our model a finite average duration of elementary avalanches
$\langle t\rangle  <\infty$ is achieved in two cases: (a) In the case
of distribution in Eq.\ (\ref{p1_0}) when the system size $L$ is finite,
hence the distribution is truncated at $t=L$; (b) In the case of
finite dissipation $c>0$, where  the distribution in Eq.\
(\ref{p1_c}) has a characteristic duration  $\xi <\infty $ for all
finite $c$ values (see Refs.\ \cite{Tetal,JT} and below).

For a finite $L$ we find a continuous flow phase (F) for low
dissipation and large driving rates, and
three regions with intermittent behavior of avalanche queues.
These are regions with subcritical (S),
nonuniversal (N), and multifractal (M) behavior, shown schematically
on the phase diagram in Fig.\ 1.
When the length separation $L\gg \xi $ holds, we find a
line in the $(r,c)$-plane where loss of particles on defects in the
interior of the pile becomes ``compensated'' by fast adding of particles
from outside.  In the region above
the compensation line the avalanche queues  exhibit a scaling behavior
with the scaling exponents depending on the driving rate: The slopes
decrease whereas the fractal dimensions increase with driving rate.
Cut-offs with a stretch-exponential behavior appear.
In the limit $c\to 0$ and when the system size $L\ll \xi $ is varied
multifractal scaling properties describe the scaled distributions,
rather than a simple finite-size scaling.

\narrowtext
\begin{figure}
\epsfxsize=82mm\epsffile[56 75 392 298]{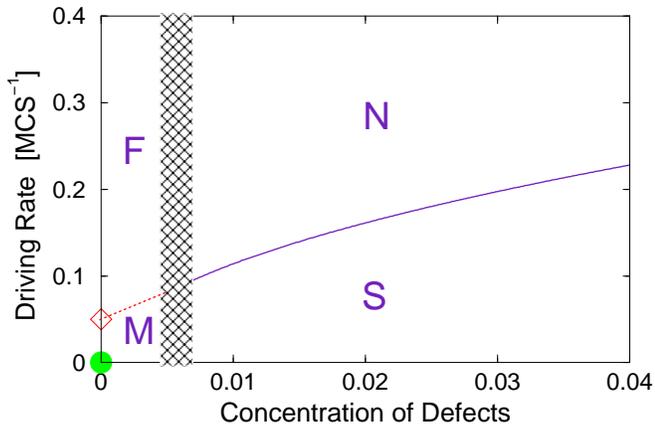}
\caption{\label{fig1}Schematic phase diagram for a finite system size
$L$. Cross-hatched area represents  a crossover region between
$\xi \ll L$ (right) and $\xi \gg L$ (left). Regions with distinct
behavior of the avalanche queues are shown: subcritical (S),
nonuniversal (N), and multifractal scaling region (M), and flow phase (F).
Solid line represents the compensation line. Transition to the flow
phase is marked by $\diamond$ for $c=0$ and by dotted line for small
$c>0$. In the origin  only single Dhar-Ramaswamy avalanches occur.}
\end{figure}

In general, the distributions of combined avalanches are characterized
by a scaling function of two arguments  of the form
\begin{equation}
P(X,r,L_c) \sim X^{-\tau _X}{\cal{P}}(XL_c^{-D_X},rL_c^{1/2}) \ ,
\label{scal-general}
\end{equation}
where $X$ represents either
duration, $t$, or number of topplings \cite{comment-n}, $n$, and $L_c
\equiv \min{(\xi ,L)}$.
The corresponding fractal dimensions $D_X$ are defined by
\begin{equation}
\langle X\rangle _{\ell} \sim \ell ^{D_X} \ ,
\label{fract-dim}
\end{equation}
where the average is taken over all combined avalanches of a selected length
$\ell $ measured along the direction of transport.

Using analogy to the queue theory and the properties of the
recurrent states we were able to derive an  exact  upper limit of the
 distributions of busy periods and to discuss the limit $L\to
\infty$. We also derived the expression for the probability of continuous
flow in the conservative limit.

The paper is organized as follows: In Sec.\ II we define the model
and consider the case of conservative dynamics by numerical
simulations on finite lattice.
In Sec.\ III we present results of simulations in the case of finite
concentration of nonconservative defects. In Sec.\ IV we present details
of the analytical results. The paper contains  a short summary of the
results and discussion in
Sec.\ V.

\section{Multifractal queues of Dhar-Ramaswamy avalanches}

The sandpile automaton model introduced by Dhar and Ramaswamy represents
an example of a self-organized criticality  with exact solution in the
limit of zero driving rate \cite{DR}. In this Section we consider the
same model at finite driving rates $0 <r\leq 1$. The dynamic rules of
the model are summarized as follows \cite{DR}: We consider a $2$-dimensional
square lattice oriented downwards, with a dynamic variable---height
$h(i,j)$---associated at each site. Grains are added at the top row only,
and mass flow is only down. The toppling
at a site $(i,j)$ occurs deterministically whenever $h(i,j) \geq h_c=2$,
and two grains are transferred downward, i.e.,
\begin{equation}
h(i,j)\to h(i,j)-2 \ ;\  h(i+1,j_\pm )\to h(i+1,j_\pm )+1 \ ,
\label{rules}
\end{equation}
where $(i+1,j_\pm )$ represents two downward  neighboring sites
to the site $(i,j)$.

The probability distribution of duration of avalanches in zero driving
rate $P_1(t) \sim t^{-\tau _t^0}{\cal{P}}(tL^{-1})$ with $\tau _t^0=3/2$
given in Eq.\ (\ref{p1_0}) becomes exact at large $t$ \cite{DR}.
In addition the dynamic exponent $z^0=1$. This implies that
the average duration $\langle t\rangle _0 \sim L^{1/2} \to \infty $.
Similarly, the area $s$ enclosed by the boundary of an avalanche is given
by the distribution  $D(s,L) \sim
s^{-\tau _s^0}{\cal{D}}(sL^{-D_s^0})$, where $\tau _s^0=4/3$ and the
fractal dimension $D_s^0=3/2$. Note that the number of toppled grains
at each active site is two, then the distribution of the number of toppled
grains within an avalanche, $D(n)$, is described by the same exponents,
i.e., $\tau _n^0=4/3$ and $D_n^0=3/2$ at zero driving rate.

A finite driving rate $r>0$ is implemented as follows. An avalanche is
started from the top and at each step of the avalanche progress a new
 particle is added with probability $r$ at a random site at first
row. We also consider a deterministic addition of particles, i.e.,
we add a particle at regular intervals $\Delta t$. Both approaches
lead to the same results when the statistics is high enough.
An added particle
may trigger a new elementary avalanche before the previous one stops,
thus making a pattern of active sites distributed over lattice. A
snapshot of growth of a combined avalanche with marked active sites is
shown in Fig.\ 2 (top).
A combined avalanche---avalanche queue---is thus determined by a
non-interrupted activity on the lattice and it stops when no more active
sites occur.  Then a new avalanche is started.
It should be noted that when $r>0$ the number of added grains is higher
than the number of combined avalanches.  Another important remark is that
the repeated toppling at
a site may occur as soon as $r>0$. This leads to the inequality
$\langle n\rangle > \langle s\rangle $, and thus $D_n >D_s$, and $z>1$.
The numerical simulations confirm these conclusions (see below).
Typically, we consider $2\times 10^6$ combined avalanches at each
driving rate and lattice size. Periodic boundary conditions are
applied in the perpendicular direction.

\begin{figure}
\epsfxsize=52mm\epsffile[188 278 421 515]{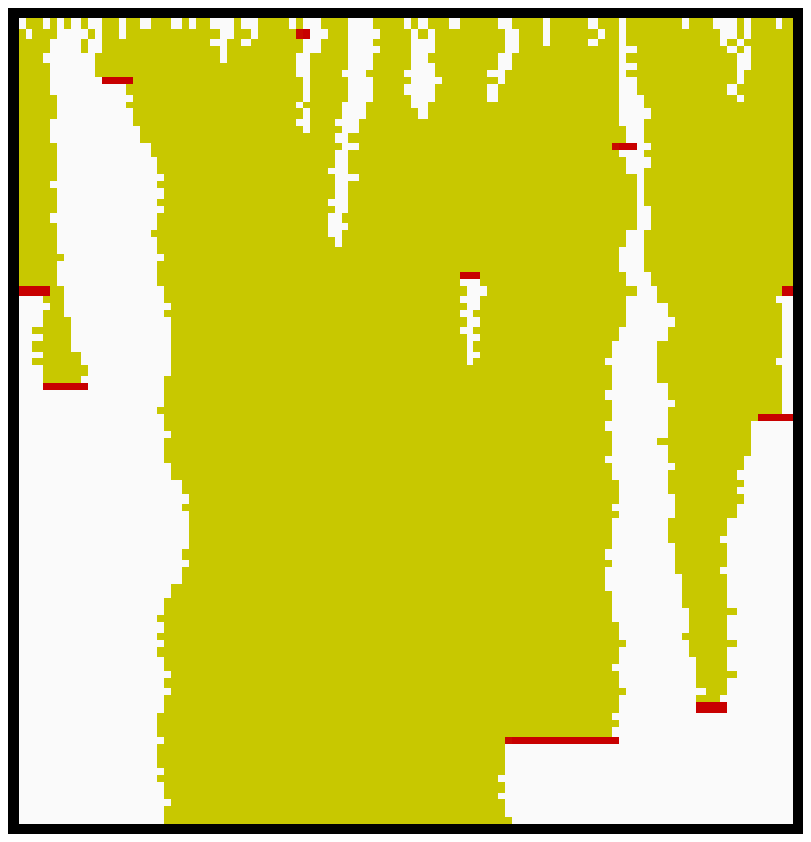}

\epsfxsize=52mm\epsffile[156 256 455 536]{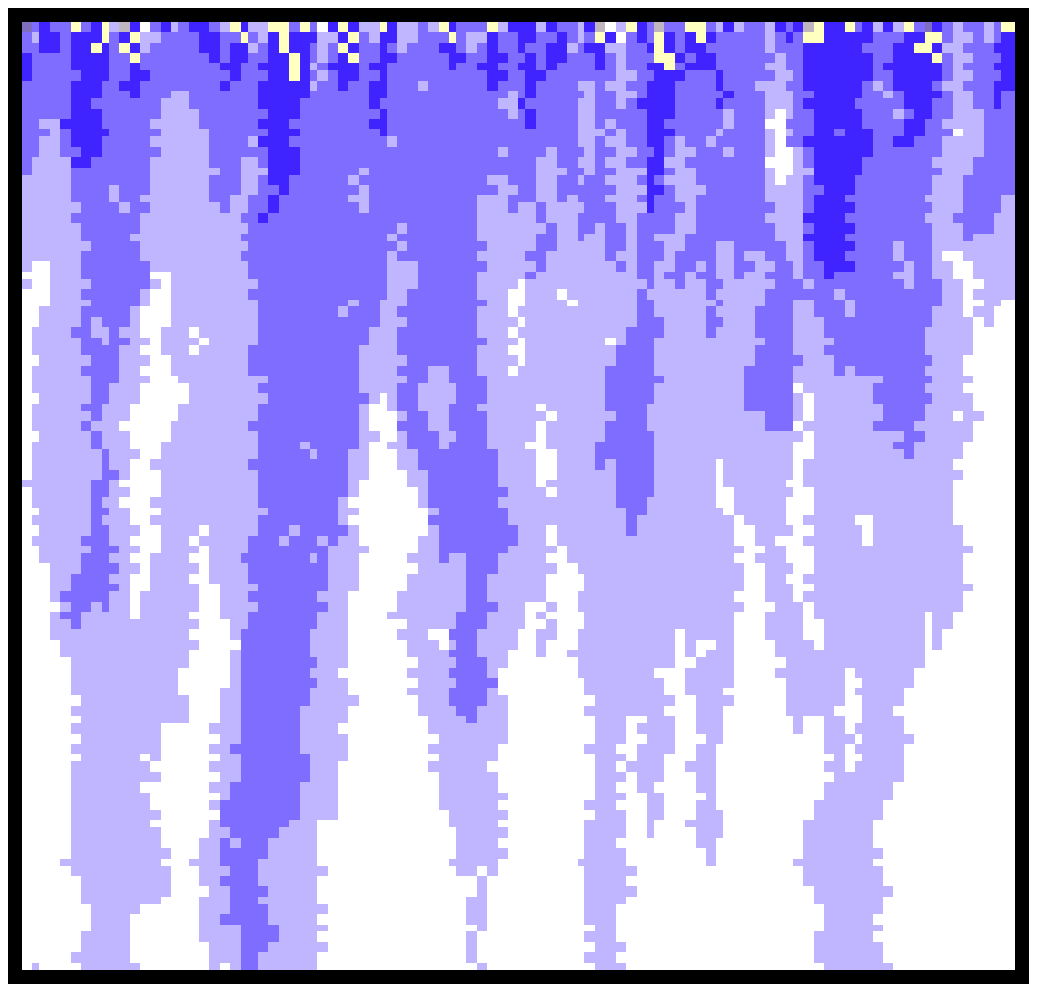}
\caption{\label{fig2} Top: Growth of a combined avalanche in the case
of conservative dynamics ($c=0$) for driving rate $r=0.05$ and $L=100$.
Nine fronts of active sites (dark) are visible.
Bottom: A complete combined avalanche in the case of dissipative dynamics with
$c=0.02$ and driving rate $r=0.5$. Different colors correspond to distinct
toppling waves.}
\end{figure}

In this Section we perform numerical simulations for  $r>0$ and
finite lattice sizes $L$. The limit $L\to \infty$ will be discussed in
Sec.\ IV. In main Figs.\ 3 and 4 are shown the integrated probability
distributions of duration $P(t^\prime \ge t,L)$ and number of topplings
$D(n^\prime \ge n,L)$ for fixed driving rate $r=0.05$ and various
lattice sizes $L$. It should be noted that both slopes and cut-offs
of these distributions appear to be different compared to ones of
the elementary avalanches. In particular, slopes decrease with $r$
(see more detailed discussion in next Section). For instance at $r=0.05$
we find $\tau _t =0.4$ and $\tau _n=0.3$, in the steep part, and
$\tau _t =0.31$ and $\tau _n=0.2$ in the flat part near the cut-off.
A cut-off in the probability
distribution of durations appear (cf. main Fig.\ 3). The characteristic
jump at $t=L$ is related to the conditional probability: an activity
lasts longer than $L$ steps only if the preceding avalanches last
not shorter than $t=L$. The jump decreases with increasing lattice size.

\begin{figure}[thb]
\epsfxsize=80mm\epsffile[43 69 508 564]{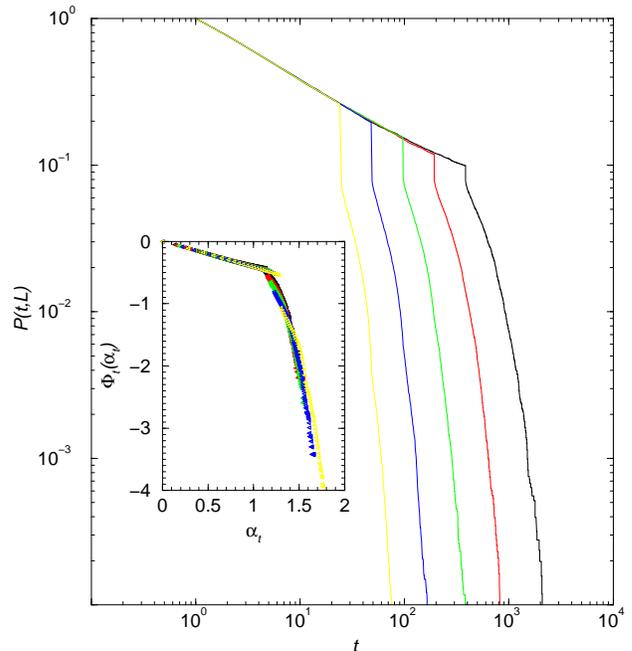}
\caption{\label{fig3}Double logarithmic plot of the integrated probability
distribution of duration (busy periods) $P(t,L)$ of queues of Dhar-Ramaswamy
avalanches vs. duration $t$, measured in Monte Carlo steps [MCS].
Fixed driving rate $r=0.05$ and various
lattice sizes $L=$24,48,96,192, and 384 are used.
Inset: Multifractal spectral function $\Phi _t(\alpha _t)$ vs. $\alpha _t$
obtained from the data in main figure according to Eq.\ (6) using
$X_0=1$, $L_0=2$.}
\end{figure}
It is interesting that these distributions can not be scaled according
to a simple finite-size scaling (with new exponents), as
one may naively expect. Instead, we find that a multifractal scaling
applies according to the law
\begin{equation}
P(X,L,r) = \left({L\over{L_0}}\right)^{\Phi _X(\alpha _X)} \ ; \
\alpha _X = {{\log(X/X_0)}\over{\log(L/L_0)}} \ ,
\label{mulifractal}
\end{equation}
where, as before, $X$ stands for $t$ or $n$. The  corresponding spectral
functions $\Phi _t(\alpha _t)$ and $\Phi _n(\alpha _n)$ are determined
numerically for $r=0.05$ and shown  in the insets to Figs.\ 3 and 4,
respectively. The spectrum depends on the driving rate. For driving
rates close to the line $r\sim L^{-1/2}$ extremely large avalanches
may appear and the scaling fits fail.
The origin of multifractality in the queues of Dhar-Ramaswamy avalanches
can be found in the fact that an unrestricted  multiple toppling at
each site may occur, and that a toppling at a given site releases a local
avalanche which propagates from that site downwards.

\begin{figure}[thb]
\epsfxsize=80mm\epsffile[43 69 508 564]{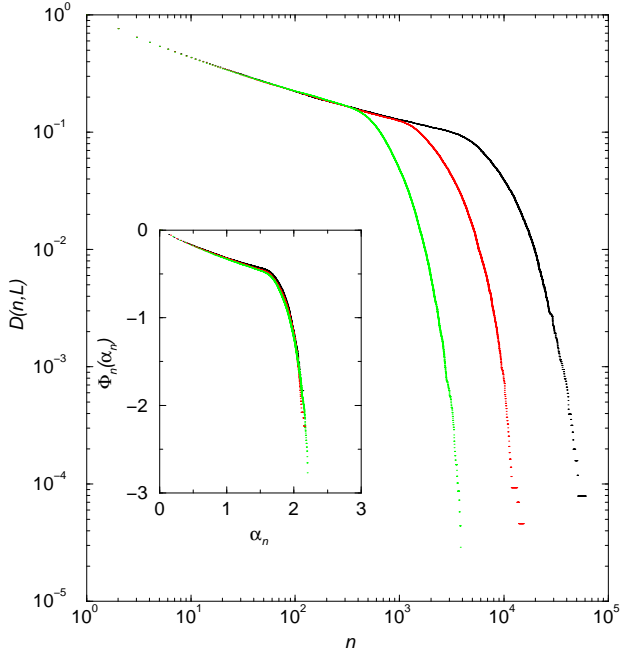}
\caption{\label{fig4}Same as Fig.\ 3 but for the integrated distribution
of number of topplings (mass) $D(n,L)$ vs. mass $n$ [number of grains].
Shown are only curves for $L=$96,192, and 384. Inset: Spectral function
$\Phi _n(\alpha _n)$ vs. $\alpha _n$. $X_0=0.97\pm 0.02$, $L_0=2.15
\pm 0.05$.}
\end{figure}

\begin{figure}[thb]
\epsfxsize=80mm\epsffile[38 73 449 298]{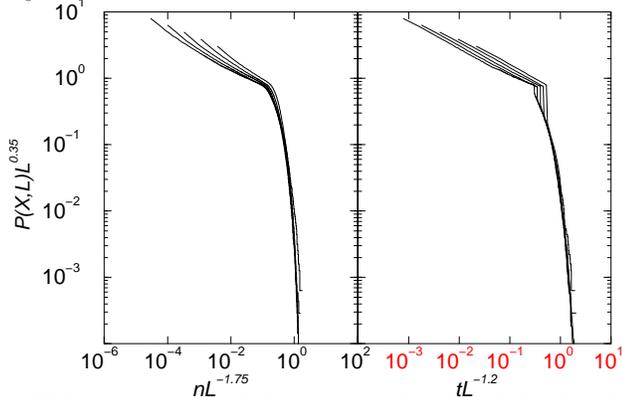}
\caption{\label{fig5} Attempted finite-size scaling fit of the data
from main Fig.\ 3 and 4 for the distribution of avalanche duration (right)
and mass (left).}
\end{figure}

For comparison we show how a finite-size scaling fit  of the
data fails. In Fig.\ 5 we present an attempt of scaling collapse of the
data shown in Figs.\ 3 and 4 above, according to the formula $P(X,L)=
L^{-\lambda }{\cal{P}}(XL^{-D_X})$.
Note that the best collapse of the tails of distributions are obtained
by fixing the fractal dimensions as $D_t=1.20$ and $D_n=1.75$, which
correspond to  $\alpha _X$ at the shoulder of
the spectrum $\Phi _t(\alpha _t)$ and  $\Phi _n(\alpha _n)$,
respectively. Fixing a smaller (larger) value for the fractal dimension
leads to systematic shifts of the distribution tails
to the right (left) with increasing $L$.
The best fit  shown in Fig.\ 5 is obtained for $\lambda =0.35$, which
satisfies (within numerical accuracy)
the scaling relation $\lambda =z(\tau_t-1) =D_n(\tau_n-1)$
with $\tau _X-1$ determined at the flat part of the distribution.
 However, as the Fig.\ 5 shows, fixing $D_X$  and  $\lambda $ leads to
the systematic shifts of the 'horizontal' part of the distributions
to the right with decreasing $L$. Fixing the exponents  independently
from the scaling relation results in crossing of the lines for
different $L$ values.

For driving rates $r> L^{-1/2}$ an ever-running avalanche may occur,
representing a continuous activity on the lattice. The flow phase
can be characterized by an average number of topplings per site,
which is expected to have a nontrivial $L$-dependence. The probability
of occurrence of the flow phase in the limit $L\to \infty $ will be
discussed in Sect.\ IV.

\section{Nonuniversality in dissipative dynamics}

In the presence of dissipative defects $c>0$ the  distribution of
elementary avalanches, which is given in Eq.\ (\ref{p1_c}), appears to
have a finite characteristic length $\xi <\infty $. Thus the average
duration at zero driving rate is finite $\langle t\rangle _0 <\infty $.
Precise value of the average duration is controlled by an external
parameter---probability of dissipation $c$, and not by system size
$L$, provided that $L> \xi$. The screening length $\xi \sim
1/c$ was first estimated numerically in Ref.\ \cite{Tetal}. A more
precise analytical expression can be derived  (see  below and Ref.\
\cite{JT}) as $\xi ^{-1} \sim -\ln (1-c)$.
Here we perform numerical simulations in the case  $L > \xi $ at
driving rates $0 < r \leq 1$.
In this range of driving rates we expect the role of lattice size
in the analysis of Sec.\ II is to be replaced by the characteristic
length $\xi $. In addition, competition  between dissipation and
driving rate leads to new phenomena.

In Fig.\ 2 (bottom) an example of a combined avalanche is shown
for $c=0.02$ and driving rate $r=0.5$. It is remarkable that the
number of topplings per site decrease with distance from the top.
Intermittency  of the dynamics as well as the occurrence of the
long-range correlations  can be seen by direct examination of
the recorded activity of the system $n(t)$ at each time step. For
the server queue the quantity $n(t)$ is interesting as the measure
of the energy which is dissipated by the server at a given moment of
time $t$. In Fig.\ 6 we show an example of the recorded signal for
certain choice of parameters  $r$, $c$, and $L$  corresponding to
the region (N) of the phase diagram (cf. Fig.\ 1).
A combined avalanche on this recording
is represented by a set of peaks between two consecutive drops of
the signal to the base line. The Fourier spectrum of the signal (shown
on top panel in Fig.\ 6) exhibits a power-law behavior. The slope
$\varphi \approx 0.9$ weakly increases  with driving rate $r$.

\begin{figure}[thb]
\epsfxsize=80mm\epsffile[43 69 508 746]{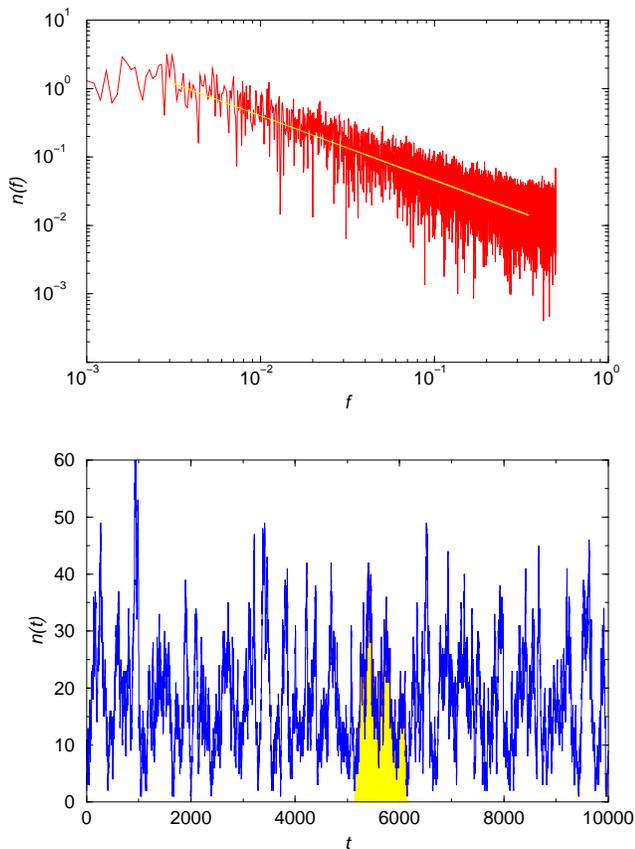}
\caption{\label{fig6} Sandpile noise $n(t)$---number of active sites at
time $t$, plotted against time $t$ [MCS] (lower panel) and its
Fourier spectrum (top panel) for fixed driving rate r=0.33.
Dissipation rate $c=0.01$ and $L=128$. Shaded area shown in the lower
panel is an example of unit signal, corresponding to a combined avalanche.}
\end{figure}

The scaling properties of the  distributions of avalanche queues
depend on the mutual ratio of the driving and dissipation rates.
In particular, the scaling function in Eq.\ (\ref{scal-general})
exhibits a nontrivial dependence on both arguments $x\equiv t\xi ^{-1}$
and $y\equiv r\xi ^{1/2}$. (Another suitable choice of variables
would be $(tc,rt^{1/2})$.)
In Fig.\ 7 we show the distribution of the avalanche mass (dissipated energy)
$n$ for fixed $c=0.01$ and various driving rates $r$. In general, the
cut-offs of the distributions increase and slopes decrease with increased
driving rates $r$. More detailed analysis of the slopes shows that for
the range of values of driving rates
the scaling behavior  of the distributions can be described by the
scaling exponents which depend on the driving rate.
The slopes of various distributions and the corresponding fractal
dimensions, which are defined in Eq.\ (\ref{fract-dim}), are
shown against driving rate in Fig.\ 8. The dissipation rate
is fixed $c=0.01$. Note that $\tau _{\ell}$ in Fig.\ 8 represents
slope of the distribution of the largest linear length reached in a
combined avalanche. For a range of values of driving rates
the scaling exponents decrease and the fractal
dimensions increase with $r$, while  the
scaling relations between various exponents are found to be satisfied
within numerical error bars.
\begin{figure}
\epsfxsize=80mm\epsffile[43 69 508 564]{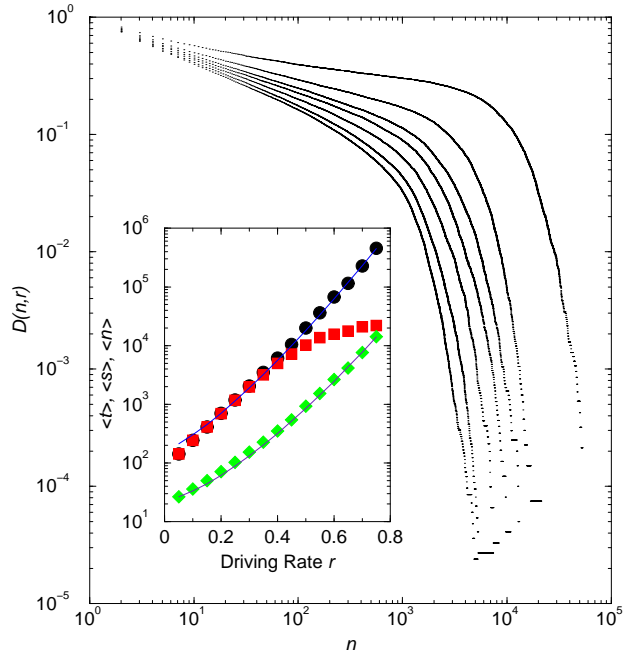}
\caption{\label{fig7}
Distribution of avalanche mass $D(n,r)$ vs. mass $n$ [number of grains]
shown in double logarithmic scale for fixed dissipation rate $c=0.01$
and $ L=192$ and for
various  driving rates $r$= 0.021,0.041,0.083,0.125,0.166,0.25, and 0.33
(left to right). Inset: Average mass (top), area (middle) and duration
(bottom) of combined avalanches plotted against driving rate [MCS$^{-1}$]
for fixed $c=0.01$ and $L=192$. At each point average is taken over
$2\times 10^6$ combined avalanches. Fitting curves: $\langle n\rangle
=167\times \exp{(11.4r^{1.28})}$, and  $\langle t\rangle =23\times
\exp{(9.6r^{1.42})}$.}
\end{figure}
The variation of the scaling exponents
can be approximated by a  linear dependence of $r$.
It is interesting to note that a qualitatively similar behavior---linear
variation of the scaling exponents with driving rate, has been measured
experimentally in the
case of Barkhausen noise in driven disordered ferromagnets \cite{BN-r}.
Our present analysis suggests that such behavior can be related
to an interplay of driving rate and dissipation at defects, and that
it applies more universally.
Dependence of the fractal dimensions (and  of the slope exponents
via scaling relations) of the avalanche queues on driving rate $r$ can
be linked to the $r$-dependence of the average length of the queue
$<N> =1+r<t>_0$.
It has been shown recently \cite{Romu} that the fractal spectrum of the
series of elementary signals in the case of transit times in the ricepile
model varies as a power of the length of series.
A precise $r$-dependence of the exponents
in the case of avalanche queues  requires additional work and will be
given elsewhere.

\begin{figure}
\epsfxsize=80mm\epsffile[52 70 449 564]{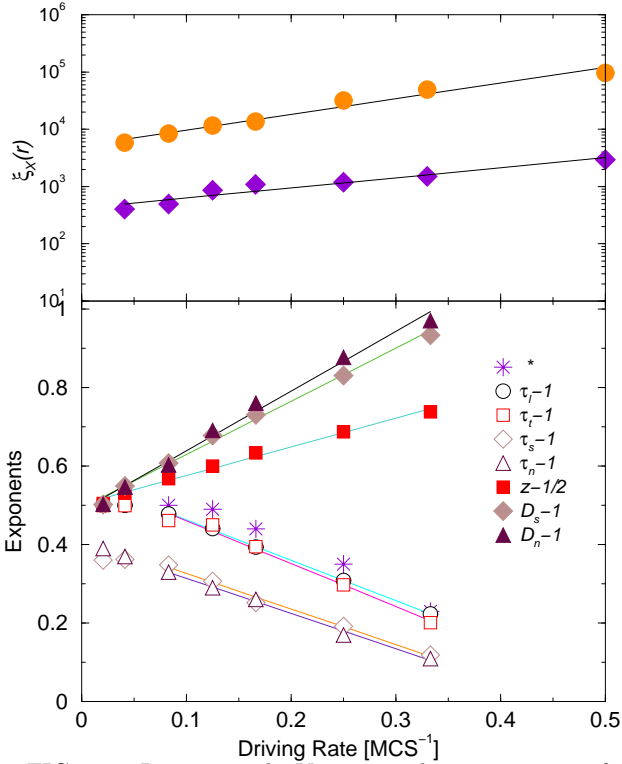}
\caption{\label{fig8} Lower panel: Various scaling exponents of the
avalanche queues plotted against driving rate for fixed dissipation
$c=0.01$ and  $L=192$.  $*$ indicates the products
$z(\tau _t-1)\approx D_n(\tau _n-1)\approx D_s(\tau_s-1) \approx
1.1(\tau _{\ell}-1)$.  Also shown are linear fits of the data.
Top panel: Amplitudes $\xi_n$ (upper curve) and $\xi_t$ defined in
Eq.\ (7) vs driving rate. Lines: fit curves satisfying
 Eq.\ (8) with $B_n=6.29\pm 0.46$ and $B_t=4.5\pm 0.51$.}
\end{figure}

The opposite effects on the exponents are obtained by increasing concentration
of defects $c$ at fixed driving rate. In particular, the slopes of the
distributions increase and fractal dimensions decrease  with {\it increasing}
concentration of defects in a limited range \cite{comment}.

An exact expression for the scaling function ${\cal{P}}(x,y)$ in Eq.\
(\ref{scal-general}) can not be guessed.
It appears that the cut of the surface ${\cal{P}}(x,y)$ at
$r=const$ for well balanced values of $c$ and $r$ can be
approximated by a stretch-exponential function, so that we have
\begin{equation}
P(X,c,r) = X^{-\tau _X(r)}\exp{(-X^{\sigma _X}/\xi _X(r))} \ .
\label{stretch-ex}
\end{equation}
Here $X$ stands for $n$ or $t$, and we find $\sigma _n =1.14 \pm 0.04$
and $\sigma _t =1.28\pm 0.06$, for the distribution of energy and duration,
 respectively. The amplitudes $\xi_X(r)$ can be fitted  (see top panel
 of Fig.\ 8)  by the following function of driving rate $r$
\begin{equation}
\xi_X(r) =  A_X(c)\exp(rB_X(c))\ ,
\label{xi-r}
\end{equation}
for a fixed dissipation rate $c$.

The observed parameter-dependence of the probability distributions  is also
reflected in the behavior of the average duration
and energy of  combined avalanches. Notice that the average duration
$\langle t\rangle $ represents the average busy period of a server in the
queue theory. Beside the average duration $\langle t\rangle $, we also
compute the average values of number of topplings (energy) $\langle
n\rangle $ and  area $\langle s\rangle $ (total number of distinct sites)
affected by the processing of a combined avalanche. These average values are
shown vs. $r$ in the inset to Fig.\ 7 for $c=0.01$. The values
$\langle t\rangle $ and $\langle n\rangle $ increase with driving rate
faster than an exponential function. Fitting the data in the inset to Fig.\ 7
we find
\begin{equation}
\langle X\rangle  = a_{0X}\exp{(a_{1X}(c)r^{\sigma })} \ ,
\label{averages}
\end{equation}
where $X\equiv n, t$ and we estimate $\sigma = 1.2 \pm 0.1$.
Note that, in contrast to durations and energies, the average area
of an avalanche is bounded
by the number of cells in the system $\langle s\rangle \leq L^2$.

As already mentioned above, for a given dissipation rate $c$ and
$L\gg \xi$ a driving rate $r_0(c)$ exists such that fast addition
of grains compensates the losses in the bulk. In fact along an extremal line
\begin{equation}
r_0(c) \sim \kappa \xi ^{-1/2} \approx  \kappa \sqrt{c} \
\label{comp-line}
\end{equation}
the coherence length  remains constant.
The existence of the compensation line  Eq.\ (\ref{comp-line}) can be
demonstrated by considering sets of data for average durations and
energies against driving rate $r$, obtained for different characteristic
length $\xi \sim 1/c$. These data can be scaled according to the
following scaling form
\begin{equation}
\langle X\rangle \xi ^{-D_X(2-\tau _X)} =
{\cal{G}}\left(r\xi ^{z(2-\tau _t)} - \kappa \right) \ .
\label{FSS-c}
\end{equation}
The corresponding scaling fits for the two cases $X\equiv t$ and
$X\equiv n$ are shown in Fig.\ 9. Notice that the respective exponents
 $D_n(2-\tau _n)=1$ and $z(2-\tau _t)=1/2$ are exact values,
thus leaving only one parameter, namely  $\kappa$, to be determined by
the fitting procedure. This is an advantage of having the exact solution
for the elementary avalanches \cite{DR}.
>From the best fit we find $\kappa =1.14 \pm 0.1$ in the given range of
values of $c$ (see caption to Fig.\ 9).
It is evident  from Fig.\ 9 that the scaling function defined in
Eq.\ (\ref{FSS-c}) increases faster than an exponential.

In the simulations  a continuous flow may occur in the  case of
dissipative dynamics at finite lattice size $L$
when the driving rate is increased. However, with increased system
size $L$ the behavior is different
from the one in the case of conservative dynamics discussed above.
>From the numerical simulations alone we can not distinguish if the
affected area  of a continuous avalanche  diverges
with the system size $L\to \infty $, or it remains finite for the range of
driving rates considered here.
We will also discuss large $L$ limit in Sec.\ IV.

We have restricted our analysis to the case where the degree of dissipation
is such that  $\xi \ll L$. For $c \to 0$, however, we have that
$\xi  \to \infty $,
thus the role of $ L$ and $\xi$ is interchanged at some finite $L$.
In the reverse limit when $L \ll \xi $ the behavior is expected to be
similar to the case of conservative dynamics at finite $L$, studied in
Sec.\ II.

\begin{figure}[thb]
\epsfxsize=78mm\epsffile[43 69 508 564]{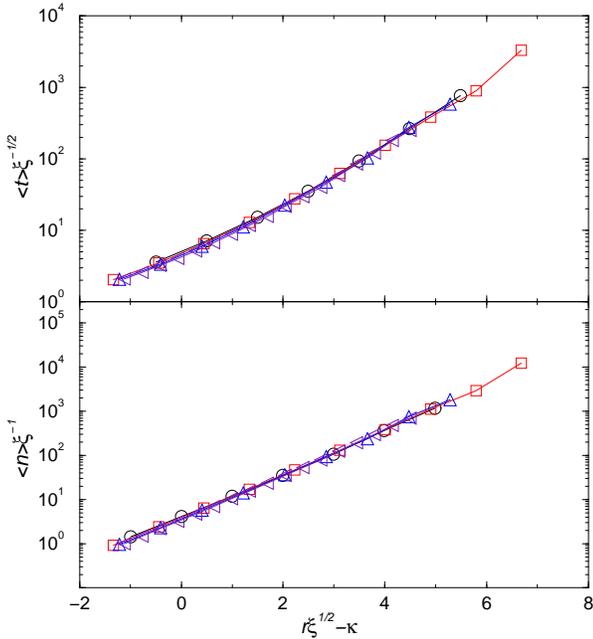}
\caption{\label{fig9} Finite-size scaling plot of the average
duration $\langle t\rangle $ (top)   and  mass
$\langle n\rangle $ (bottom) according to Eq.\ (\ref{FSS-c}).
Data are taken for values of $c$=0.01,0.0125,0.015, and 0.02
such that $\xi  <L=192$ is satisfied.
Note that the exponents are exact and the fitted value
$\kappa = 1.14 $ within numerical error bars.}
\end{figure}

\section{Analytical  results}

We start the analytical description of the model with the
assumption that individual avalanches, which form each combined
avalanche, are statistically independent. By the definition of the model,
each toppling in a single avalanche occurs later than those in the
previous avalanches, so that the individual avalanches never intersect
in the space-time points. Nevertheless, the next avalanche is sensitive
to the configuration of occupation numbers left by the previous avalanches.
In this way the individual avalanches are dependent on preceding avalanches.
On the other hand, it is known from Abelian properties
of the directed sandpile model \cite{DR}, that the recurrent state is
characterized by the independent distribution of occupation numbers
zero and one at each site. Hence, one can expect that this property
of recurrent state provides independent distribution
of single avalanches at least for asymptotically large systems.
This assumption allows us to consider  the process of driving of
the directed sandpile automaton  as a sequence of independent identically
distributed (i.i.d.) events.  Another important consequence of the
statistical independence is that  stops of combined
avalanches can be considered as recurrent events, i.e., the probability of
two successive stops $Prob(t_1,t_2)$ at the moments $t_1$ and $t_1+t_2$ is
given by the product $Prob(t_1)Prob(t_2)$.

We  consider the
probability distribution $F(t)$ that a combined avalanche
starting at the moment $t=0$  stops {\it for the first time}
 at the discrete moment $t$. This means that $F(t)$ is the probability
 that the  stop of all  preceding elementary avalanches  occurs
till the moment $t$. Thus, $F(t)$ coincides with the probability
distribution of duration of combined avalanches, when an ensemble of
events is considered.
 Along with $F(t)$, it is useful to consider the function $U(t)$ which
is defined as the probability that all preceding single avalanches stop
to the moment $t$ {\it regardless} of  how many stops of
combined avalanches occurred  before $t$.
Noting that stops of combined avalanches are recurrent events, we can
write for $F(t)$ and $U(t)$ the following identity \cite{Feller}
\begin{equation}
U(t)=F(1)U(t-1)+F(2)U(t-2)+...+F(t)U(0)
\label{2}
\end{equation}
where it is convenient to put F(0)=0 and U(0)=1. For the generating
functions defined by
\begin{equation}
\label{3}
u(s)=\sum_{t=0}^{\infty} s^t U(t)
\end{equation}
and
\begin{equation}
\label{4}
f(s)=\sum_{t=1}^{\infty} s^t F(t)
\end{equation}
one  easily gets from Eq. (\ref{2}) the known equation of the theory of
recurrent events \cite{Feller}
\begin{equation}
\label{5}
f(s)=\frac{u(s)-1}{u(s)} \ .
\end{equation}
The total  probability that a combined avalanche ever stops is
given by $f(1)$.
Therefore, the probability that a combined avalanche never stops,
i.e., the probability of a continuous flow, which is given by
\begin{equation}
\label{6}
F(\infty)=1-f(1),
\end{equation}
does not vanish if $u(1)$ in Eq.\ (\ref{5}) is finite $u(1)<\infty$  .

\subsection{Case $c=0$}

In the case of conservative dynamics ($c=0$) the
 probability distribution of durations of elementary  avalanches
is given by Eq.\ (\ref{p1_0}).
Then we can estimate the probability $U(t)$ as follows.
Let $\Delta t \equiv 1/r$ be the average
time interval between addition of successive  particles to the first row.
The probability $Prob(x \leq t)$ that a single avalanche has duration
less than $t$ is
\begin{equation}
\label{8}
Prob(x \leq t) \sim 1- \frac{b}{t^{1/2}}
\end{equation}
for large $t$, where $b$ is a constant of the order unity when $L$
is large. Then for times $t \gg \Delta t$ we have
\begin{equation}
\label{9}
U(t) \sim \left(1- \frac{b}{t^{1/2}}\right)\left(1-
\frac{b}{(t-\Delta t)^{1/2}}\right)\cdots \left(1-
\frac{b}{(\Delta t)^{1/2}}\right) \ .
\end{equation}
Introducing $k=t/\Delta t$ we can write Eq.\ (\ref{9}) as the sum
\begin{equation}
\label{10}
\ln U(t) = \sum_{n=1}^{k}\ln \left(1-\frac{b}{(n \Delta t)^{1/2}}\right)
\leq - \frac{k^{1/2}}{(\Delta t)^{1/2}} \ .
\end{equation}
 Approximating the sum  by an integral leads to
 \begin{equation}
\label{11}
U(t)\sim \left({{\sqrt{rt}-b\sqrt{r}}\over{1-
b\sqrt{r}}}\right)^{-b^2r}\exp{[-2b(r\sqrt{t}-\sqrt{r})]} \
\end{equation}
for $t\gg 1$ and $0 < r \leq 1$.
For the infinite lattice  there exists a  constant $c_2$ such that
\begin{equation}
\label{12}
u(1) \leq c_2 \sum_{t=0}^{\infty} \exp (- r\sqrt{t}) < \infty \ .
\end{equation}
Therefore, we can find that for all finite driving rates $r>0$
there is a non-zero probability of continuous flow. In particular,
the sum in Eq. (\ref{12}) diverges at small $r$ as
\begin{equation}
\label{13}
u(1) \sim \frac{1}{r^2}
\end{equation}
leading to the probability of stop $f(1) \sim 1-2r^2b^2 $, which
 decreases from unity as soon as a finite driving rate is applied. Then the
probability of continuous flow  Eq.\ (\ref{5}) increases from zero by the
same amount, i.e.,
\begin{equation}
\label{14}
F(\infty) \sim r^2,  r \to  0 \ .
\end{equation}
For large $r$ we expect that
\begin{equation}
\label{15}
1 - F(\infty)\sim \exp (-r) \ .
\end{equation}
If the size of the system is finite ($L<\infty $) the probability of
stops $U(t)$ is bounded from {\it above} by a finite value
\begin{equation}
\label{16}
U(t) \le (rL)^{-b^2r/2}\exp (-2brL^{1/2} )\ ,
\end{equation}
which follows from Eq.\ (\ref{11}) taking only the dominant $t$-behavior
for $t=L \gg 1$.

\subsection{Case $c>0$}

In the case of finite dissipation rate $c>0$  the dissipation
leads to a finite characteristic length $\xi $ in the
distribution of elementary avalanches in Eq.\ (\ref{p1_c}).
This can be  easily demonstrated using  mean-field arguments
\cite{Ivash.Priez} in reaction-diffusion systems.
Let us suppose that at each site of the lattice
one of species $A$ or $B$ is living. These species represent two
possible states of the original model: $A$ corresponds to the empty
site, $B$ to the occupied site.
Due to the external driving force, new particles $\phi$ are added
to the first row of the lattice at rate $r$. The
propagation of particles can be described by the following rules
\begin{equation}
\label{18}
A+\phi \rightarrow B,\  B+\phi \rightarrow A+2\phi \ .
\end{equation}
The kinetic equations corresponding to this scheme of ``chemical'' reactions
are
\begin{equation}
\label{19}
\dot{n}_A(\ell )=n_{\phi}(\ell )[n_B(\ell )-n_A(\ell )]
\end{equation}
\begin{equation}
\label{20}
\dot{n}_B(\ell )=n_{\phi}(\ell )[n_A(\ell )-n_B(\ell )]
\end{equation}
\begin{equation}
\label{21}
\dot{n}_{\phi}(\ell )=-n_{\phi}(\ell )+2n_{\phi}(\ell -1)n_B(\ell )(1-c)
\end{equation}
where $n_A(\ell )$, $n_B(\ell )$, and $n_{\phi}(\ell )$ are concentrations
of species  $A$, $B$ and $\phi$, respectively, at the $\ell $-th row.
In the steady state we have
$\dot{n}_A=\dot{n}_B=\dot{n}_{\phi}=0$ and Eqs.\ (\ref{19})-(\ref{21})
lead to the simple conditions for the concentrations
$n_A=n_B=1/2$ and \cite{comment-addition}
\begin{equation}
\label{23}
n_{\phi}(\ell )=r(1-c)^\ell \ .
\end{equation}
For $c>0$, the density of particles $\phi$ and, hence, the number of
topplings in an avalanche decay exponentially with the distance $\ell $
from the top as $n_{\phi}(\ell )=r\exp(-\ell/\xi )$. This implies that
the characteristic length of the avalanche
is $\xi ^{-1} \sim -\ln(1-c) \sim c$.
Therefore the above  results, in particular Eq.\ (\ref{16}), obtained for
the case of finite lattices $L$ and $c=0$
apply also  for the case $c>0$ by the substitution $L\to L_c$ with
\begin{equation}
\label{24}
L_c =\min\{\xi ,L \} \ .
\end{equation}

\subsection{Bounds for the busy time}

If the combined avalanches are finite (i.e., there is no continuous flow),
we can estimate
their average duration using the known theorem from the theory of recurrent
events (Ref.\ \cite{Feller}, Ch. XIII, Theorem 3 ). According to this theorem,
the inverse average time of combined avalanches $\langle t \rangle ^{-1}$
coincides with the limit of the sequence $U(t)$ when $t \rightarrow \infty$.
Using the  bound for $U(t)$ given by Eq.\ (\ref{16}), we get
\begin{equation}
\label{29}
\langle t \rangle \geq (rL)^{b^2r/2}\exp (2brL^{1/2} )
\end{equation}
The true asymptotics of $\langle t \rangle$ possibly contains an
additional  prefactor $L^{1/2}$ like in the case for $r=0$. Notice that
we get numerically that the average duration as a function of $r$ (cf.
inset to Fig.\ 7) increases faster than
the exponential, which agrees with Eq.\ (\ref{29}).

The combination $rL^{1/2}$ appears as  a characteristic parameter
determining the duration of combined avalanches. Thus, for $\xi < L$ it
follows from Eqs.\ (\ref{29}) and (\ref{23}) that the coherence length
remains constant if $r$ varies with $c$ as  $r \sim \sqrt{c}$, i.e. the
increasing driving rate compensates the dissipation.

Another interesting feature of the probability distributions at finite
driving rates is the occurrence of a stretch-exponential cutoffs both
in dissipative and non-dissipative case (cf. Figs.\ 3,4,7).
Indeed, we can see from Eq.\ (\ref{2}) that $U(t)>F(t)$ for all finite $t$.
Therefore, for non-dissipative case we have an exponential decay of
combined avalanches
\begin{equation}
\label{25}
P(t) \sim F(t) < (rt)^{-b^2r/2}\exp(-2br\sqrt{t})
\end{equation}
in the thermodynamic limit $L \rightarrow \infty$, which follows directly
from Eq.\ (\ref{11}) for large $t$.

For finite lattice sizes $L$ or finite dissipation $c>0$ the function
$U(t)$ is bounded from above by a constant given in Eq.\ (\ref{16})
with $L$ replaced by $L_c$. In this case we can find the
origin of an exponential cut-off in the following way.
Consider an enveloping process which corresponds to propagation of the
front of combined avalanches. Duration of an
 elementary avalanche starting at $t_{i0}$ is a simple linear function
of time, $t_i =t-t_{i0}$. The enveloping process consists of those
topplings  which occur at the maximal distance from the time axis
at each moment of time $t$. The position $x$
of the front is an one-dimensional random walk confined to the interval
$[0,\xi ]$. Starting from $x=0$, the walk performs a step ahead with some
probability, and a step back, the length of which is a random variable.
The combined avalanche stops if the random walk returns to the origin
where it is trapped. The probability of the return to the origin from
an arbitrary position $x$ is not smaller than $U(\xi )$, where
\begin{equation}
\label{26}
U(\xi ) = (r\xi )^{-b^2r/2}\exp(-2br\xi ^{1/2}) \ ,
\end{equation}
and $\xi \sim 1/c$ as above.
The survival time $t$ of the random walk under consideration does not
exceed the period of successful tests in the Bernoulli scheme with the
probability of ``success'' $1-U(\xi )$. The period of tests in the
Bernoulli process has the geometrical distribution
\begin{equation}
\label{27}
Prob(x=t)=(1-U(\xi ))^t U(\xi )\ .
\end{equation}
Hence, the time distribution for combined avalanches is bounded from
above by the exponential function in Eq.\ (\ref{27}). Using Eq.\
(\ref{26}), we obtain
\begin{equation}
\label{28}
P(t) \sim F(t)< A(1-U(\xi ))^t \sim A \exp(-tU(\xi ))\ ,
\end{equation}
where $A$ is a constant and $U(\xi )$ is given by Eq.\ (\ref{26}).
Note that this expression represents an upper bound of the distribution
of busy periods in Eq.\ (\ref{stretch-ex}). Therefore $1/U(\xi )$ plays
the role of an effective correlation length at finite $r$, in a qualitative
agreement with the numerical data and Eq.\ (\ref{xi-r}) of previous Section.

\section{Conclusions and discussion}

We have shown that a finite driving rate $r$ is a relevant perturbation,
which alters self-organized critical states in the directed sandpile
automata. In the case of conservative dynamics $r$ couples to $\langle
t\rangle _0 \sim L^{1/2}$, thus  leading to enhanced effects when the
length scale is increased $L\to \infty$. A continuous flow eventually
occurs, in which critical long-range correlations are destroyed.
On a finite length scale, $L_c =\min \{\xi ,L\} <\infty $
either due to finite screening length $\xi $ or finite system size $L$,
the critical states  occur with qualitatively new correlation properties,
which is manifested in (i) a multifractal scaling of combined avalanches
when $L \ll \xi $, and (ii) occurrence of compensation between driving
and dissipation  along a line $r_0(c)\sim \xi ^{-1/2}\sim \sqrt{c}$,
when $L\gg \xi $.
How precisely the effective coherence length $\xi_{eff}(r)$ of combined
avalanches increases with driving rate depends on details of the
relaxation process and grain addition. In the case of a finite
input current at each site of the system, we find a finite toppling rate at
all scales, $n_{\phi }(\ell )\sim r/c$, compatible with $\xi _{eff}(r)
\to \infty$. However, if grains
are  added only at the top, the correlation length increases
exponentially with $r$ in the range $0<r\le 1$, but  remains finite
presumably up to large driving rates. Here we restricted the
discussion to the case $r\leq 1$, where queues of Dhar-Ramaswamy
avalanches  occur. Avalanche queuing for this range of driving rates
is peculiar to our model, due to strictly local critical height rule
and the directed transport. In the rice-pile and in the symmetric
Abelian models \cite{Al-Maya,BTW-r} a  perfect queuing is prevented by
the collision of avalanches, which occurs at any finite $r>0$.
Owing to the exact solution for behavior of elementary avalanches \cite{DR},
we were able to study  properties of the queue in detail. In particular,
we find that average busy period is bounded from below by an exponential
function $\langle t\rangle \geq (rL)^{b^2r/2}\exp{(2brL^{1/2})}$. The
avalanche queue
can be regarded as a multifractal set, in which  the average length  is
regulated by the driving rate as $\langle N\rangle =1+rL^{1/2}$.
There are no waiting times for elementary avalanches, therefore
our model corresponds to the realization  known  in the queue theory as
``infinite number of servers''. Hence, the average number of jobs $q_n$ that
can be served in parallel is unlimited.  It should be stressed that our
cellular automaton  represents a new example in the queue theory in which
queuing jobs are distributed according to a power-law distribution with
the exponent $\nu <2$ and average duration of jobs is limited by a
control  parameter $L_c=\min \{\xi ,L \}$. We hope that more practical
examples of this class can be found.
We also believe that the study of the scaling properties of the queues,
as we have done in this work,  adds a new aspect  which has not been
considered so far in the queue theory.
The observed nonuniversal scaling properties of avalanche queues can be
related to variation of the average length of the queue with driving
rate.  The scaling exponents are found to vary approximately
linearly with the driving rate $r$. A similar $r$-dependence was
observed experimentally in other driven disordered  systems, and seems
to apply more generally.

A continuous activity on the lattice, corresponding to a flow phase
occurs for $r\geq  L^{-1/2}$ in the case of conservative
dynamics. On an infinite lattice $L\to \infty $ probability of continuous
flow increases from zero as $F(\infty )\sim 2b^2r^2$, whereas probability
of an intermittent avalanche-like flow decreases from unity with the same
rate $f(1)\sim 1- 2b^2r^2$. For the case of dissipative dynamics $c>0$
an effective coherence length increases with $r$ for $1/c < L\to \infty$.
 Our results suggest that the  probability $f(1)$ remains finite even
at high driving rates. In the limit $L\to \infty $ the compensation line
extends to the point $c\to 0$, $r\to 0$.
For the range of driving rates studied in this work  we expect that the
transport properties of grains in this model remain
unchanged (looked at the time scale of avalanche propagation), compared
to the transport at zero driving rate \cite{BT-unp}. Collision of
avalanches, which occurs first at rates $r>1$ may accelerate the grain
transport, possibly resulting in a  new scaling behavior of the
distribution of transit times.   Notice that due to local critical height
rules and deterministic topplings the depth of the active zone (defined
in Ref.\ \cite{Al-Maya}) does not change in the flow phase of our model.

The  analytical results in Sec.\ IV are
derived assuming that elementary avalanches may be considered as
independent events. We checked by computing  numerically  correlation
function between events in a  queue for finite $L$ that
rather weak correlations occur. The correlations increase  with
the ``distance'' $\tau $ between avalanches as $\tau ^\eta $, where
$\eta =0.05\pm 0.01$ with statistical error bars.

\acknowledgments
Work of B.T. was  supported by the Ministry
of Science and Technology of the Republic of Slovenia.
Work of V.P. was supported in part by the International
Slovenian-Russian project. B.T. wishes to thank A. Corral for
discussions.

\end{multicols}


\begin{references}
\bibitem{comment-n}Number of topplings $n$ occurring during an
avalanche (also called avalanche mass) in this model
is proportional to the energy
dissipated in the avalanche. Here we use all three terms equally.
\bibitem{comment-addition}Note that the behavior is different
in the case of addition of particles at {\it each site} on the lattice,
when we have a constant number of topplings at
all distances $n_{\phi}(\ell ) = (r/c)[1+(1-c)^{\ell -1}]$ controlled by the
ratio $r/c$.
\end{references}
\end{document}